\documentclass{kapproc} 

\usepackage{procps} 

\usepackage[dvips]{graphicx}

\upperandlowercase

\kluwerbib 

\def    \beq       {\begin{equation}}
\def    \eeq       {\end{equation}}

\def    \Angstrom  {\,{\rm\AA}}
\def    \micron    {\,\mu{\rm m}}
\def    \cm     {\,{\rm cm}}

\def	\g	{\,{\rm g}}

\def    \H      {{\rm H}}

\def    \K      {\,{\rm K}}
\def    \mag    {\,{\rm mag}}

\def    \ltsim  {\lesssim}      
\def    \gtsim  {\gtrsim}       

\def	\mum	{\,\mu{\rm m}}
\def	\um	{\,\mu{\rm m}}
\def	\simali	{\sim\,}

\def	\ppm    {\,{\rm ppm}}

\def    \msun   {{\rm M_\odot}}

\def    \magni  {\,{\rm mag}}
\def	  \xsun         {\left[{\rm X/H}\right]_{\odot}}
\def	  \csun         {\left[{\rm C/H}\right]_{\odot}}

\def	  \osun         {\left[{\rm O/H}\right]_{\odot}}
\def	  \fesun        {\left[{\rm Fe/H}\right]_{\odot}}
\def	  \mgsun        {\left[{\rm Mg/H}\right]_{\odot}}
\def	  \sisun        {\left[{\rm Si/H}\right]_{\odot}}
\def	  \xdust        {\left[{\rm X/H}\right]_{\rm dust}}
\def	  \cdust        {\left[{\rm C/H}\right]_{\rm dust}}

\def	  \odust        {\left[{\rm O/H}\right]_{\rm dust}}
\def	  \fedust       {\left[{\rm Fe/H}\right]_{\rm dust}}
\def	  \mgdust       {\left[{\rm Mg/H}\right]_{\rm dust}}
\def	  \sidust       {\left[{\rm Si/H}\right]_{\rm dust}}
\def	  \xgas        {\left[{\rm X/H}\right]_{\rm gas}}
\def	  \cgas        {\left[{\rm C/H}\right]_{\rm gas}}

\def	  \ogas        {\left[{\rm O/H}\right]_{\rm gas}}
\def	  \fegas       {\left[{\rm Fe/H}\right]_{\rm gas}}
\def	  \mggas       {\left[{\rm Mg/H}\right]_{\rm gas}}
\def	  \sigas       {\left[{\rm Si/H}\right]_{\rm gas}}
\def	  \mux         {\mu_{\rm X}}

\def	  \muo         {\mu_{\rm O}}

\def	  \muh         {\mu_{\rm H}}
\def       \lambdamax    {\lambda_{\rm  max}}
\def       \Pmax         {P_{\rm  max}}

\def	  \Urad		{U}

\def	  \dN	        {dN_{\rm H}}
\def      \WCD          {{\rm\bf WCD04}}
\def      \BG           {{\rm\bf BG96}}
\def    \ltsim  {\lesssim}      
\def    \gtsim  {\gtrsim}       
\def    \simlt  {\lesssim}      
\def    \simgt  {\gtrsim}       
\newcommand{\figwidth}{3.8in}
\newcommand{\figwidthland}{3.5in}
\newcommand{\figwidthL}{4.4in}

\def\lesssim{\mathrel{\hbox{\rlap{\hbox{\lower4pt\hbox{$\sim$}}}\hbox{$<$}}}}
\def\gtrsim{\mathrel{\hbox{\rlap{\hbox{\lower4pt\hbox{$\sim$}}}\hbox{$>$}}}}

\begin{document}

\articletitle{\large The Warm, Cold and Very Cold Dusty Universe}
\vskip -0.25in
\author{Aigen Li}

\affil{Theoretical Astrophysics Program, University of Arizona,
       Tucson, AZ 85721}

\vspace*{-11.0em}
{\noindent \normalsize\tt invited review article for 
the international symposium {\bf ``Penetrating Bars 
Through Masks of Cosmic Dust''} (Johannesburg, South Africa, 
6--12 June 2004), edited by D.L. Block \& K. Freeman, 
Springer, pp.\,535--559}
\vspace*{6.1em}

\vskip -0.18in
\begin{abstract}
We are living in a dusty universe: 
dust is ubiquitously seen in a wide variety 
of astrophysical environments, ranging from 
circumstellar envelopes around cool red giants 
to supernova ejecta, from diffuse and dense 
interstellar clouds and star-forming regions 
to debris disks around main-sequence stars, 
from comets to interplanetary space 
to distant galaxies and quasars. 

These grains, spanning a wide 
range of sizes from a few angstroms to a few micrometers,
play a vital role in the evolution of galaxies as an absorber,
scatterer, and emitter of electromagnetic radiation,
as a driver for the mass loss of evolved stars,
as an essential participant in the star and planet
formation process, as an efficient catalyst for
the formation of H$_2$ and other simple molecules
as well as complex organic molecules which may lead to
the origins of life, as a photoelectric heating agent 
for the interstellar gas, and as an agent shaping
the spectral appearance of dusty systems such as 
protostars, young stellar objects, evolved stars
and galaxies.

In this review I focus on the dust grains in the space
between stars (interstellar dust), with particular emphasis
on the extinction (absorption plus scattering) and emission
properties of {\it cold} submicron-sized ``classical'' grains 
which, in thermal equilibrium with the ambient interstellar 
radiation field, obtain a steady-state temperature of 
$\simali$15--25$\K$, {\it warm} nano-sized (or smaller) 
``ultrasmall'' grains which are, upon absorption of 
an energetic photon, transiently heated to 
temperatures as high as a few hundred to over 1000$\K$,
and the possible existence of a population of {\it very cold}
($<$10\,K) dust. Whether dust grains can really get down to 
``temperature'' less than the 2.7$\K$ cosmic microwave 
background radiation temperature will also be discussed.
The robustness of the silicate-graphite-PAHs interstellar
dust model is demonstrated by showing that the infrared 
emission predicted from this model closely matches that
observed for the Milky Way, the Small Magellanic Cloud,
and the ringed Sb galaxy NGC\,7331.  
\end{abstract}

\vskip -0.06in
\begin{keywords}
Interstellar dust -- extinction -- absorption -- infrared emission
\end{keywords}

\section{\large\hspace{-8mm} Introduction}
The space between the stars (interstellar space) 
of the Milky Way Galaxy and other galaxies is filled 
with gaseous ions, atoms, molecules (interstellar gas) 
and tiny dust grains (interstellar dust).
The first direct evidence which pointed to 
the existence of interstellar gas came from 
the ground-based detection of Na and Ca$^+$ 
optical absorption lines (Hartmann 1904).\footnote{%
  Struve \& Elvey (1938) appear to be first to show 
  that the interstellar gas is mostly hydrogen. 
  }   
This did not gain wide acceptance until Struve (1929) 
showed that the strength of the Ca$^+$ K-line was 
correlated with the distance of the star. 
The true interstellar nature of this gas 
was further supported by the detection of 
the first interstellar molecules CH, CH$^+$ 
and CN (Swings \& Rosenfeld 1937, McKellar 1940, 
Douglas \& Herzberg 1941).
The presence of dark, obscuring matter in 
the Milky Way Galaxy was also recognized at 
the beginning of the 20th century (e.g. see Barnard 1919).
Trumpler (1930) first convincingly showed that 
this ``obscuring matter'' which dims and reddens starlight
consists of small solid dust grains.
The dust and gas are generally well mixed in 
the interstellar medium (ISM), as demonstrated
observationally by the reasonably uniform correlation 
in the diffuse ISM between the hydrogen column density
$N_{\rm H}$ and the dust extinction color excess 
or reddening $E(B-V)\equiv A_B-A_V$: 
$N_{\rm H}/E(B-V) \approx 5.8\times 10^{21}\mag^{-1}\cm^{-2}$
(Bohlin et al. 1978), where $A_B$ and $A_V$ are 
the extinction at the $B$ ($\lambda$\,=\,4400$\Angstrom$)
and $V$ ($\lambda$\,=\,5500$\Angstrom$) wavelength bands.
From this correlation one can estimate the gas-to-dust 
ratio to be $\simali$210 in the diffuse ISM.\footnote{%
 Let interstellar grains be approximated by a single
 size of $a$ (spherical radius) with a column density
 of $N_{\rm d}$. The gas-to-dust mass ratio is
 \begin{equation}
 \frac{m_{\rm gas}}{m_{\rm dust}}
 \approx \frac{\left(\mu_{\rm H}+
 \left[{\rm He/H}\right]_{\rm ISM}\,\mu_{\rm He}\right)\,N_{\rm H}}
 {\left(4/3\right)\pi a^3 \rho_{\rm d}N_{\rm d}}
 \approx \frac{1.4\mu_{\rm H}\,N_{\rm H}}
 {\left(4/3\right)\pi a^3 \rho_{\rm d}N_{\rm d}} 
 \end{equation}
 where $\mu_{\rm H}$ and $\mu_{\rm He}$ are respectively
 the atomic weight of H and He; 
 $\left[{\rm He/H}\right]_{\rm ISM}$ is the interstellar
 He abundance (relative to H), which is taken to be that
 of the solar value, $\simali$0.1; $\rho_{\rm d}$ is the 
 mass density of interstellar dust. 
 The hydrogen (of all forms)-to-dust column density 
 $N_{\rm H}/N_{\rm d}$ can be derived from 
 \begin{equation}
 \frac{N_{\rm H}}{E(B-V)} = \frac{N_{\rm H}}{A_V/R_V}
 = \frac{R_V\,N_{\rm H}}{1.086\,\pi a^2 Q_{\rm ext}(V)\,N_{\rm d}}
 \end{equation}
 where $R_V$ is the total-to-selective extinction ratio,
 $Q_{\rm ext}(V)$ is the dust extinction efficiency at 
 $V$-band ($\lambda=5500$\,\AA). Therefore, the gas-to-dust
 ratio can be readily estimated from
 \begin{equation}
 \frac{m_{\rm gas}}{m_{\rm dust}}
 \approx 1.14\,\mu_{\rm H}\,\frac{Q_{\rm ext}(V)}{a\rho_{\rm d}}
 \frac{1}{R_V} \frac{N_{\rm H}}{E(B-V)}
 \approx 212\,\left(\frac{Q_{\rm ext}[V]}{1.5}\right)
 \left(\frac{0.1\mum}{a}\right)
 \left(\frac{2.5\g\cm^{-3}}{\rho_{\rm d}}\right) ~~.
 \end{equation}
 If we take canonical numbers of $R_V$$\approx$3.1,
 $Q_{\rm ext}(V)$$\approx$1.5,
 $a$=0.1$\mum$, $\rho_{\rm d}$$\approx$2.5$\g\cm^{-3}$,
 $m_{\rm gas}/m_{\rm dust}$ would be around 210.
 }
Despite its tiny contribution to the total mass 
of a galaxy,\footnote{%
 In our Milky Way Galaxy, interstellar matter 
 (gas and dust; $7\pm3\,\times 10^{9}\msun$),
 contributes roughly $\simali$20\% of 
 the total stellar mass ($4\pm2\,\times 10^{10}\msun$). 
 Therefore, the mass fraction of interstellar dust
 is just $\simali$0.1\% in our Galaxy
 ($\simali$$1\pm0.2\,\times 10^{11}\msun$ within 10\,kpc,
  Kennicutt 2001)!
 }
interstellar dust has a dramatic effect on the physical
conditions and processes taking place within the universe,
in particular, the evolution of galaxies and the formation
of stars and planetary systems (see the introduction section 
of Li \& Greenberg 2003).

In this review I will concentrate on the radiative
properties of interstellar dust. I will distinguish
the dust in terms of 3 components:
{\bf cold dust} with steady-state temperatures 
of 15$\K$$\simlt$$T$$\simlt$25$\K$ in thermal
equilibrium with the solar neighbourhood interstellar 
radiation field, {\bf very cold dust} with $T$$<$10$\K$
(typically $T$$\simali$5$\K$), and {\bf warm dust} undergoing
``temperature spikes'' via single-photon heating.
I will first summarize in \S2 
the observational constraints on the physical 
and chemical properties of interstellar dust. 
I will then discuss in \S3 the heating and cooling 
properties of the warm and cold interstellar dust components.
In \S4 I will demonstrate the robustness of 
the silicate-graphite-PAHs interstellar grain model
by showing that this model closely reproduces 
the observed infrared (IR) emission of the Milky Way,
the Small Magellanic Cloud, and the ringed Sb galaxy
NGC\,7331. The possible existence of a population of 
very cold dust (with $T$$<$10$\K$) in interstellar space 
will be discussed in \S5. Whether ultrasmall grains
can really cool down to $T$$<$2.7$\K$ and appear in
absorption against the cosmic microwave background (CMB)
radiation will be discussed in \S6. 

\vspace{-2mm}
\section{\large\hspace{-8mm} Observational Constraints on Interstellar Dust}
\vspace{-1mm}
Our knowledge of interstellar dust regarding its composition, 
size and shape is mainly derived from its interaction with 
electromagnetic radiation: attenuation (absorption and scattering)
and polarization of starlight, and emission of IR and 
far-IR radiation. The principal observational keys, both direct
and indirect, used to constrain the properties of dust are 
the following:
\begin{itemize}
\item {\bf (1) \underline{Interstellar Extinction}.}~~
Extinction is a combined effect of absorption and scattering:
a grain in the line of sight between a distant star and 
the observer reduces the starlight by a combination of 
scattering and absorption (the absorbed energy is then 
re-radiated in the IR and far-IR).
\vskip -0.10in
\begin{itemize}
\vskip -0.15in
\item The wavelength dependence of interstellar extinction
-- ``interstellar extinction curve'', 
most commonly determined from the ``pair-method'',\footnote{%
  In this method, the wavelength dependence of interstellar 
  extinction is obtained by comparing the spectra of
  two stars of the same spectral type, one of which 
  is reddened and the other unreddened.  
  }
rise from the near-IR to the near-UV, 
with a broad absorption feature at about 
$\lambda^{-1}$$\approx$4.6$\mum^{-1}$ 
($\lambda$$\approx$2175$\Angstrom$),
followed by a steep rise into the far-UV
$\lambda^{-1}$$\approx$10$\mum^{-1}$. 
$\longrightarrow$
{\bf The extinction curve tells us the size 
(and to a less extent, the composition) 
of interstellar dust.}
Since it is generally true that a grain
absorbs and scatters light most effectively 
at wavelengths comparable to its size 
$\lambda$$\approx$$2\pi a$ (Kr\"ugel 2003), 
there must exist in the ISM a population 
of large grains with $a$$\simgt$$\lambda/2\pi$$\approx$$0.1\mum$
to account for the extinction at visible wavelengths,
and a population of ultrasmall grains 
with $a$$\simlt$$\lambda/2\pi$$\approx$$0.016\mum$
to account for the far-UV extinction at 
$\lambda$=$0.1\mum$ (see Li 2004a for details).
\item The optical/UV extinction curves show
considerable regional variations.
$\longrightarrow$ {\bf Dust grains on different 
sightlines have different size distributions 
(and/or different compositions).}
\item The optical/UV extinction curve in the wavelength 
range of 0.125\\$\le$$\lambda$$\le$$3.5\mum$
can be approximated by an analytical formula involving 
only one free parameter: $R_V$$\equiv$$A_V/E(B-V)$, 
the total-to-selective extinction ratio
(Cardelli, Clayton, \& Mathis 1989),
wher-\\eas the near-IR extinction curve 
(0.9$\mum$$\le$$\lambda$$\le$$3.5\um$) can 
be fitted reasonably well by a power law 
$A(\lambda)$$\sim$$\lambda^{-1.7}$, 
showing little environmental variations. 

The Galactic mean extinction curve
is characterized by $R_V\approx 3.1$. 
Values of $R_V$ as small as 2.1 
(the high latitude translucent molecular cloud HD 210121; 
Larson, Whittet, \& Hough 1996; Li \& Greenberg 1998) 
and as large as 5.6 (the HD 36982 molecular cloud 
in the Orion nebula) have been observed 
in the Galactic regions.
More extreme extinction curves have been 
reported for extragalactic objects.\footnote{%
  Falco et al.\ (1999) found $R_V$$\approx$1.5 for 
  an elliptical lensing galaxy at $z_l$$\approx$0.96, 
  and $R_V$$\approx$ 7.2 for a spiral lensing galaxy 
  at $z_l$$\approx$0.68.
  Wang et al.\ (2004) found 
  the extinction curves for two intervening quasar 
  absorption systems at $z$$\approx$1.5 to have 
  $R_V$$\approx$0.7,1.9.
  } 
$\longrightarrow$ 
{\bf The optical/UV extinction curve and the value of $R_V$ 
depend on the environment}: lower-density regions have 
a smaller $R_V$, a stronger 2175$\Angstrom$ hump and 
a steeper far-UV rise ($\lambda^{-1}$$>$4$\mum^{-1}$),
implying smaller grains in these regions;
denser regions have a larger $R_V$, a weaker 2175$\Angstrom$
hump and a flatter far-UV rise, implying larger grains. 
\item In the Small Magellanic Cloud (SMC), the extinction curves
of most sightlines display a nearly linear steep rise with 
$\lambda^{-1}$ and an extremely weak or absent 2175$\Angstrom$ 
hump (Lequeux et al.\ 1982; Pr\'evot et al.\ 1984). 
$\longrightarrow$ {\bf Grains in the SMC are smaller than
those in the Milky Way diffuse ISM as a result of either
more efficient dust destruction in the SMC due to its harsh
environment of the copious star formation associated with
the SMC Bar or lack of growth due to the low-metallicity of 
the SMC, or both.} Regional variations also exist 
in the SMC extinction curves.\footnote{%
 For example, there is at least one line of sight 
 (Sk 143$=$AvZ 456) with an extinction curve with 
 a strong 2175\AA\ hump detected (Lequeux et al.\ 1982; 
 Pr\'evot et al.\ 1984; Bouchet et al.\ 1985; 
 Thompson et al.\ 1988; Gordon \& Clayton 1998). 
 This sightline passes through the SMC wing, a region 
 with much weaker star formation (Gordon \& Clayton 1998).
 The sight lines which show no 2175\AA\ hump all pass
 through the SMC Bar regions of active star formation 
 (Pr\'evot et al.\ 1984; Gordon \& Clayton 1998). 
 }
\item The Large Magellanic Cloud (LMC) extinction curve 
is characterized by a weaker 2175$\Angstrom$ hump
and a stronger far-UV rise than the average 
Galactic extinction curve (Nandy et al.\ 1981;
Koornneef \& Code 1981).\footnote{%
  Strong regional variations in extinction
  properties have also been found 
  in the LMC (Clayton \& Martin 1985; 
  Fitzpatrick 1985,1986; Misselt, Clayton, \& Gordon 1999):
  the sightlines toward the stars inside or near 
  the supergiant shell, LMC 2, which lies on the southeast
  side of the 30 Dor star-forming region,
  have very weak 2175$\Angstrom$ hump (Misselt et al.\ 1999).  
  }
\end{itemize}

\vspace{-3mm}
\item {\bf (2) \underline{Interstellar Polarization}.}~~
For a non-spherical grain, the light of distant stars 
is polarized as a result of differential extinction for 
different alignments of the electric vector of the radiation. 
\vskip -0.12in
\begin{itemize}
\vskip -0.15in
\item The interstellar polarization curve rises from the IR, 
has a maximum somewhere in the optical and then decreases 
toward the UV. $\longrightarrow$ {\bf This tells us that 
(1) some fraction of interstellar grains are both non-spherical
and aligned by some process; (2) the bulk of the aligned grains
responsible for the peak polarization 
(at $\lambda$$\approx$0.55$\mum$) have typical sizes of
$a$$\approx$$\lambda/2\pi$$\approx$0.1$\mum$;
and (3) the ultrasmall grain component responsible for 
the far-UV extinction rise is either spherical or not aligned.} 
\item The optical/UV polarization curve $P(\lambda)$ can 
be closely approximated by the ``Serkowski law'',
an empirical function: 
$P(\lambda)/\Pmax\\ = \exp [-K\ln^2(\lambda/\lambdamax)]$,
where the only one free parameter, $\lambdamax$,
is the wavelength where the maximum polarization 
$\Pmax$ occurs, and $K$ is the width parameter: 
$K$$\approx$$1.66 \lambdamax$\,+\,0.01;
$\lambdamax$ is indicative of grain size and
correlated with $R_V$: $R_V \approx (5.6\pm 0.3)\lambdamax$ 
($\lambdamax$ is in micron; see Whittet 2003).
$\longrightarrow$ {\bf The sightlines with larger
$\lambdamax$ are rich in larger grains and have 
larger $R_V$ for their extinction curves.} 
A close correlation between $\lambdamax$ and the size
of the aligned grains [e.g., $\lambdamax$$\approx$$2\pi a (n-1)$
for dielectric cylinders of radius $a$ and refractive index $n$] 
is predicted by interstellar extinction calculations. 
\item The near-IR (1.64$\mum$$<$$\lambda$$<$$5\mum$) 
polarization is better approximated by 
a power law $P(\lambda)$\,$\propto$\,$\lambda^{-\beta}$,
with $\beta$$\simeq$$1.8\pm0.2$, 
independent of $\lambda_{\rm max}$
(Martin \& Whittet 1990, Martin et al.\ 1992).
\end{itemize}

\vspace{-3mm}
\item {\bf (3) \underline{Interstellar Scattering}.}~~
Scattering of starlight by interstellar dust is revealed
by reflection nebulae (dense clouds illuminated by embedded
or nearby bright stars), dark clouds (illuminated by the general
interstellar radiation field [ISRF]), and the diffuse Galactic 
light (``DGL''; starlight scattered off the general diffuse 
ISM of the Milky Way Galaxy illuminated by the general ISRF). 
The scattering properties of dust grains 
(albedo = ratio of scattering to extinction,
and phase function) provide a means of 
constraining the optical properties of the grains
and are therefore indicators of their size and composition
and provide diagnostic tests for dust models.
\vskip -0.15in
\begin{itemize}
\vskip -0.15in
\item The albedo in the near-IR and optical
is quite high ($\simali$0.6), with a clear dip
to $\simali$0.4 around the 2175$\Angstrom$ hump,
a rise to $\simali$0.8 around $\lambda^{-1}$$\approx$6.6$\mum^{-1}$,
and a drop to $\simali$0.3 by $\lambda^{-1}$$\approx$10$\mum^{-1}$;
the scattering asymmetry factor almost monotonically rises from
$\simali$0.6 to $\simali$0.8 from $\lambda^{-1}$$\approx$1$\mum^{-1}$
to $\lambda^{-1}$$\approx$10$\mum^{-1}$ (see Gordon 2004).
$\longrightarrow$ {\bf An appreciable fraction of
the extinction in the near-IR and optical must arise from
scattering; the 2175$\Angstrom$ hump is an absorption feature
with no scattered component (see \S2.4); and ultrasmall grains
are predominantly absorptive.} 
\item Surprisingly high near-IR albedo has been reported
for several regions: $\simali$0.86 at the K$^{\prime}$ band
($\lambda$$\approx$2.1$\mum$) for the prominent dust lane
in the evil eye galaxy NGC\,4826 (Witt et al.\ 1994), 
$\simali$0.9 at the K$^{\prime}$ band for the dust in 
the M\,51 arm (Block 1996), 
and $\simali$0.7 at the J ($\lambda$$\approx$1.26$\mum$)
and H ($\lambda$$\approx$1.66$\mum$) bands and 
$\simali$0.6 at the K ($\lambda$$\approx$2.16$\mum$)
band for the Thumbprint Nebula (Lehtinen \& Mattila 1996),
in comparison with $\simali$0.2 at $\lambda$=2.2$\mum$
predicted by the conventional dust models 
(Draine \& Lee 1984; Li \& Greenberg 1997).
$\longrightarrow$ This implies that for NGC\,4826 and M\,51
(1) a population of grains at least $\simali$0.5$\mum$ in 
radii, twice as large as assumed by standard models, may exist 
in these environments and are responsible for the near-IR scattering; 
and/or (2) the measured high K$^{\prime}$ surface-brightness 
and the deduced high albedo may in part be caused by the thermal 
continuum emission from stochastically heated ultrasmall grains
(Witt et al.\ 1994; Block 1996).
For the Thumbprint Nebula, the high near-IR albedo is readily
explained by grain growth (larger-than-average grain sizes)
and the accretion of an ice mantle (see Pendleton, Tielens,
\& Werner 1990; \S8 in Li \& Greenberg 1997).
\item Scatterings of X-rays by interstellar dust have also been 
observed as evidenced by ``X-ray halos'' formed around an 
X-ray point source by small-angle scattering. The intensity 
and radial profile of the halo depends on the composition, 
size and morphology and the spatial distribution of 
the scattering dust particles (see Dwek et al.\ 2004 for
a review). The total and differential cross sections
for X-ray scattering approximately vary as $a^4$ and $a^6$,
respectively, {\bf the shape and intensity of X-ray halos 
surrounding X-ray point sources therefore provide one of 
the most sensitive constraints on the largest grains 
along the sightline,} while these grains are gray at optical 
wavelengths and therefore the near-IR to far-UV extinction 
modeling is unable to constrain their existence.

A recent study of the X-ray halo around Nova Cygni 1992 by 
Witt, Smith, \& Dwek (2001) pointed to the requirement of 
large interstellar grains ($a$$\simali$0.25--2$\mum$),
consistent with the recent Ulysses and Galileo detections
of interstellar dust entering our solar system 
(Gr\"{u}n et al.\ 1994; Frisch et al.\ 1999; Landgraf et al.\ 2000). 
But Draine \& Tan (2003) found that the silicate-graphite-PAH
model with the dust size distributions derived from the near-IR 
to far-UV extinction modeling (Weingartner \& Draine 2001a)
and IR emission modeling (Li \& Draine 2001b) is able to
explain the observed X-ray halo.
\end{itemize}

\item {\bf (4) Spectroscopic Extinction and Polarization Features: 
\underline{The 2175$\Angstrom$}\\\underline{Extinction Hump}~
-- the strongest spectroscopic extinction feature.}
\vskip -0.10in
\begin{itemize}
\vskip -0.10in
\item Its strength and width vary with 
environment while its peak position is quite invariant:
the central wavelength of this feature varies by only
$\pm$0.46\% (2$\sigma$) around 2175$\Angstrom$ 
(4.6$\mum^{-1}$), while its FWHM varies by $\pm$12\% 
(2$\sigma$) around 469$\Angstrom$ ($\approx$1$\mum^{-1}$).
\item Its carrier remains unidentified 39 years after 
its first detection (Stecher 1965). It is generally
believed to be caused by aromatic carbonaceous 
(graphitic) materials, very likely a cosmic mixture
of polycyclic aromatic hydrocarbon (PAH) molecules
(Joblin, L\'eger \& Martin 1992; Li \& Draine 2001b;
Draine 2003a).
\item For most sightlines, this feature is unpolarized.
So far only two lines of sight toward {\small HD\,147933} 
and {\small HD\,197770} have a weak 2175$\Angstrom$ polarization 
feature detected (Clayton et al.\ 1992; Anderson et al.\ 1996; 
Wolff et al.\ 1997; Martin, Clayton, \& Wolff 1999). 
Even for these sightlines, the degree of alignment 
and/or polarizing ability of the carrier should be very small 
(see Li \& Greenberg 2003).
\item Except for the detection of scattering in the 
2175$\Angstrom$ hump in two reflection nebulae (Witt, Bohlin, 
\& Stecher 1986), the 2175$\Angstrom$ hump is thought to be
predominantly due to absorption, suggesting its carrier 
is sufficiently small to be in the Rayleigh limit.
\item The detections of this feature in distant objects have
been reported by Malhotra (1997) in the composite absorption
spectrum of 96 intervening Mg\,II absorption systems at
0.2$<$$z$$<$2.2; by Cohen et al.\ (1999) in a damped Ly$\alpha$
absorber (DLA) at $z$\,=\,0.94; by Toft, Hjorth \& Burud (2000) 
in a lensing galaxy at $z$\,=\,0.44;
by Motta et al.\ (2002) in a lensing galaxy at $z$\,=\,0.83;
and very recently by Wang et al.\ (2004) in 3 intervening
quasar absorption systems at 1.4$\simlt$$z$$\simlt$1.5.      
\end{itemize}

\item {\bf (5) Spectroscopic Extinction and Polarization Features: 
\underline{The 9.7$\um$}\\ \underline{and 18$\um$ (Silicate) 
Absorption Features}~ -- the strongest IR Absorption features.}
\vskip -0.10in
\begin{itemize}
\vskip -0.10in
\item Ubiquitously seen in a wide range of astrophysical 
environments, these features are almost certainly due to
silicate minerals: they are respectively ascribed to 
the Si-O stretching and O-Si-O bending modes in some form 
of silicate material 
{\small (e.g. olivine Mg$_{2x}$Fe$_{2-2x}$SiO$_4$).}
\item The observed interstellar silicate bands are broad and 
relatively featureless. $\longrightarrow$ {\bf Interstellar 
silicates are largely amorphous rat-\\her than crystalline.}
Li \& Draine (2001a) estimated that the {\small amount} of 
$a$$<$1$\um$ crystalline silicate grains in the diffuse ISM 
is $<$5\% of the solar Si abundance.\footnote{%
 Kemper, Vriend \& Tielens (2004) found that crystalline 
 fraction of the interstellar silicates along the sightline 
 towards the Galactic Center is $\simali$0.2\%.
 }
\item The strength of the 9.7$\mum$ feature is approximately
$\Delta \tau_{9.7\mum}/A_V$$\approx$\\$1/18.5$ in the local
diffuse ISM. $\longrightarrow$ {\bf Almost all Si atoms have 
been locked up in silicate dust, if assuming solar abundance 
for the ISM.}\footnote{%
 The amount of Si (relative to H) required to deplete in dust
 to account for the observed 9.7$\mum$ feature strength is
 \begin{equation}
 \left[\frac{\rm Si}{\rm H}\right] = 
 \frac{\Delta \tau_{9.7\mum}}{N_{\rm H}}
 \frac{1}{\kappa_{\rm sil}^{\rm abs}(9.7\mum)\,\mu_{\rm sil}}
 = \frac{\Delta \tau_{9.7\mum}}{A_V}
   \frac{A_V}{N_{\rm H}} 
   \frac{1}{\kappa_{\rm sil}^{\rm abs}(9.7\mum)\,\mu_{\rm sil}}
 \end{equation}
 where $\kappa_{\rm sil}^{\rm abs}(9.7\mum)$ is the silicate
 mass absorption coefficient at $\lambda$=9.7$\mum$;
 $\mu_{\rm sil}$ is the silicate molecular weight.
 With $\kappa_{\rm sil}^{\rm abs}(9.7\mum)$$\approx$2850$\cm^2\g^{-1}$
 and $\mu_{\rm sil}$$\approx$172$\mu_{\rm H}$ for amorphous
 olivine MgFeSiO$_4$, the local diffuse ISM
 ($\Delta \tau_{9.7\mum}/A_V$$\approx$1/18.5,
  $A_V/N_{\rm H}$$\approx$$5.3\times$$10^{-22}\magni\cm^2$)
 requires $\left[\frac{\rm Si}{\rm H}\right]$$\approx$35$\ppm$.
 }
\item The 9.7 and 18$\mum$ silicate absorption features 
are polarized in some interstellar regions, most of which
are featureless. The only exception is AFGL 2591, a molecular 
cloud surrounding a young stellar object, which displays
a narrow feature at 11.2$\um$ superimposed on the broad 
9.7$\mum$ polarization band, generally attributed to 
annealed silicates (Aitken et al.\ 1988).
\end{itemize}

\item {\bf (6) Spectroscopic Extinction and Polarization Features: 
\underline{The 3.4$\um$}\\ \underline{(Aliphatic Hydrocarbon) 
Absorption Feature.}}
\vskip -0.10in
\begin{itemize}
\vskip -0.15in
\item Widely seen in the diffuse ISM of the Milky Way Galaxy
and other galaxies (e.g. Seyfert galaxies and ultraluminous
infrared galaxies, see Pendleton 2004 for a recent review),
this strong absorption band is attributed to the C-H stretching 
mode in {\bf aliphatic hydrocarbon dust}. Its exact nature remains 
uncertain, despite 23 years' extensive investigation with over 20 
different candidates proposed (see Pendleton \& Allamandola 2002 
for a summary). So far, the experimental spectra of hydrogenated
amorphous carbon (HAC; Schnaiter, Henning \& Mutschke 1999,
Mennella et al.\ 1999) and the organic refractory residue,
synthesized from UV photoprocessing of interstellar ice mixtures
(Greenberg et al.\ 1995), 
provide the best fit to both the overall feature
and the positions and relative strengths of the 3.42$\um$, 
3.48$\um$, and 3.51$\um$ subfeatures corresponding to symmetric 
and asymmetric stretches of C--H bonds in CH$_2$ and CH$_3$ groups. 
Pendleton \& Allamandola (2002) attributed this feature to
hydrocarbons with a mixed aromatic and aliphatic character. 
\item The 3.4$\mum$ band strength for interstellar aliphatic
hydrocarbon dust is unknown. If we adopt a mass absorption
coefficient of 
$\kappa_{\rm abs}(3.4\mum)$$\simali$1500$\cm^2\g^{-1}$
(Li \& Greenberg 2002), we would require $\simali$68\,ppm C 
to be locked up in this dust component to account for 
the local ISM 3.4$\mum$ feature 
($\Delta\tau_{3.4\mum}/A_V$$\approx$1/250; Pendleton et al.\ 1994). 
\item {\bf This feature is ubiquitously seen in the diffuse ISM
while never detected in molecular clouds}.
Mennella et al.\ (2001) and Mu\~{n}oz Caro et al.\ (2001) 
argue that this can be explained by the competition between 
dehydrogenation (destruction of C-H bonds by UV photons) 
and rehydrogenation (formation of C-H bonds by H atoms interacting 
with the carbon dust): in diffuse clouds, rehydrogenation prevails 
over dehydrogenation; in dense molecular clouds, dehydrogenation 
prevails over rehydrogenation as a result of the reduced amount of 
H atoms and the presence of ice mantles which inhibits the
hydrogenation of the underlying carbon dust by H atoms while
dehydrogenation can still proceed since the UV radiation can
penetrate the ice layers.
\item Whether the origin of the interstellar aliphatic hydrocarbon 
dust occurs in the outflow of carbon stars or in the ISM itself
is a subject of debate. The former gains strength from the close
similarity between the 3.4$\um$ interstellar feature 
and that of a carbon-rich protoplanetary nebula CRL 618 
(Lequeux \& Jourdain de Muizon 1990; Chiar et al.\ 1998). 
However, the survival of the stellar-origin dust in the diffuse 
ISM is questionable (see Draine 1990). 
\item So far, {\bf no polarization has been detected for this
feature} (Adamson et al.\ 1999).\footnote{%
 Hough et al.\ (1996) reported the 
 detection of a weak 3.47$\um$ polarization feature 
 in the Becklin-Neugebauer object in the OMC-1 Orion 
 dense molecular cloud, attributed to carbonaceous 
 materials with diamond-like structure.
 See Li (2004a) and Jones \& d'Hendecourt (2004)
 for a detailed discussion on interstellar diamond.
 }
Spectropolarimetric measurements for 
both the 9.7$\mum$ silicate 
and the 3.4$\mum$ hydrocarbon features 
for the same sightline would allow a direct test of 
the silicate core-hydrocarbon mantle interstellar dust model
(Li \& Greenberg 1997), since this model predicts that 
the 3.4$\mum$ feature would be polarized if the 9.7$\mum$
feature (for the same sightline) is polarized 
(Li \& Greenberg 2002).
\end{itemize}

\item {\bf (7) Spectroscopic Extinction and Polarization Features: 
\underline{The Ice Ab-}\\\underline{sorption Features.}}
\vskip -0.10in
\begin{itemize}
\vskip -0.15in
\item Grains in dark molecular clouds 
(usually with $A_V$$>$3$\magni$) obtain ice mantles
consisting of H$_2$O, NH$_3$, CO,
CH$_3$OH, CO$_2$, CH$_4$, H$_2$CO 
and other molecules (with H$_2$O as the dominant species),
as revealed by the detection of various ice absorption 
features (e.g., H$_2$O: 3.1, 6.0$\um$; CO: 4.67$\um$; 
CO$_2$: 4.27, 15.2$\um$; CH$_3$OH: 3.54, 9.75$\um$; 
NH$_3$: 2.97$\um$; CH$_4$: 7.68$\um$;
H$_2$CO: 5.81$\um$; OCN$^{-}$: 4.62$\um$;
see Boogert \& Ehrenfreund 2004 for a review).
\item Polarization has been detected in the 3.1$\um$ 
H$_2$O, the 4.67$\um$ CO and 4.62$\um$ OCN$^{-}$
absorption features (e.g. see Chrysostomou et al.\ 1996).
\end{itemize}

\item {\bf (8) \underline{The Extended Red Emission: 
Dust Photoluminescence.}}
The ``Extended Red Emission'' (ERE), widely seen
in reflection nebulae, planetary nebulae, HII regions, 
the Milky Way diffuse ISM, and other galaxies, 
is a far-red continuum emission in excess of what 
is expected from simple scattering of starlight 
by interstellar dust (see Witt \& Vijh 2004).
The ERE is characterized by a broad, featureless band 
between $\simali$5400$\Angstrom$ and 9500$\Angstrom$,
with a width $600\Angstrom \ltsim {\rm FWHM} \ltsim 1000\Angstrom$
and a peak of maximum emission at 
$6100\Angstrom \ltsim \lambda_{\rm p} \ltsim 8200\Angstrom$, 
depending on the physical conditions of the environment 
where the ERE is produced. 
\vskip -0.10in
\begin{itemize}
\vskip -0.15in
\item The ERE is generally attributed to photoluminescence (PL) 
by some component of interstellar dust,
powered by UV/visible photons with a photon conversion 
efficiency $\eta_{\rm PL}$$\gg$10\% 
(Gordon, Witt, \& Friedmann 1998).
$\longrightarrow$
{\bf The ERE carriers are very likely in the nanometer size range} 
because nanoparticles are expected to luminesce efficiently 
through the recombination of the electron-hole pair
created upon absorption of an energetic photon,
since in such small systems the excited electron 
is spatially confined and the radiationless transitions 
that are facilitated by Auger and defect related recombination 
are reduced (see Li 2004a).
\item The ERE carrier remains unidentified. Various candidate
materials have been proposed, but most of them appear unable
to match the observed ERE spectra and satisfy 
the high-$\eta_{\rm PL}$ requirement 
(Draine 2003a; Li \& Draine 2002a; Li 2004a; Witt \& Vijh 2004).
Promising candidates include PAHs (d'Hendecourt et al.\ 1986)
and silicon nanoparticles (Ledoux et al.\ 1998, Witt et al.\ 1998,
Smith \& Witt 2002), but both have their own problems
(see Li \& Draine 2002a).
\end{itemize}

\item {\bf (9) Spectroscopic Emission Features: 
\underline{The 3.3, 6.2, 7.7, 8.6, 11.3$\mum$}\\
\underline{``Unidentified Infrared (UIR) Emission features.}}
The distinctive set of ``UIR'' emission features at
3.3, 6.2, 7.7, 8.6, and 11.3$\mum$ are seen in a wide
variety of objects, including planetary nebulae, 
protoplanetary nebulae, reflection nebulae, 
HII regions, photodissociation fronts, 
circumstellar envelopes, and external galaxies. 
\vskip -0.10in
\begin{itemize}
\vskip -0.15in
\item {\bf These ``UIR'' emission features are now generally 
identified as vibrational modes of PAHs}
(L\'{e}ger \& Puget 1984; Allamandola, Tielens, \& Barker 1985):
C--H stretching mode (3.3$\mum$), 
C--C stretching modes (6.2 and 7.7$\mum$),
C--H in-plane bending mode (8.6$\mum$),
and C--H out-of-plane bending mode (11.3$\mum$).\footnote{%
 Other C--H out-of-plane bending modes 
 at 11.9, 12.7 and 13.6$\mum$ have also been detected.
 The wavelengths of the C--H out-of-plane bending modes 
 depend on the number of neighboring H atoms:
 11.3$\mum$ for solo-CH (no adjacent H atom),
 11.9$\mum$ for duet-CH (2 adjacent H atoms),
 12.7$\mum$ for trio-CH (3 adjacent H atoms),
 and 13.6$\mum$ for quartet-CH (4 adjacent H atoms).
 }
{\bf The relative strengths and precise wavelengths 
of these features are dependent on the PAH size
and its ionization state} which is controlled by
the starlight intensity, electron density, and
gas temperature of the environment
(Bakes \& Tielens 1994, Weingartner \& Draine 2001b,
Draine \& Li 2001). 
\item {\bf Stochastically heated by the absorption of 
a single UV/visible photon (Draine \& Li 2001; Li 2004a), 
PAHs, containing {\small $\simali$45$\ppm$} C, 
account for $\simali$20\% of the total power 
emitted by interstellar dust in the Milky Way 
diffuse ISM (Li \& Draine 2001b).}
\item {\bf The excitation of PAHs does not require UV
photons;} long wavelength (red and far-red) photons are 
also able to heat PAHs to high temperatures so that they
emit efficiently at the UIR bands. 
This is because {\bf the PAH electronic absorption edge 
shifts to longer wavelengths upon ionization and/or 
as the PAH size increases}.\footnote{%
  Li \& Draine (2002b) have modeled the excitation 
  of PAH molecules in UV-poor regions. 
  It was shown that the astronomical PAH model provides
  a satisfactory fit to the UIR spectrum of vdB\,133, 
  a reflection nebulae with the lowest ratio of UV to total
  radiation among reflection nebulae with detected UIR band 
  emission (Uchida, Sellgren, \& Werner 1998).  
  }
\item No polarization has been detected 
for the PAH emission features
(Sellgren, Rouan, \& L\'{e}ger 1988). 
\item The {\bf PAH absorption features} at 3.3$\um$ 
and 6.2$\um$ have been detected in both local sources 
and Galactic Center sources (Schutte et al.\ 1998;
Chiar et al.\ 2000). The strengths of these features
are in good agreement with those predicted from
the astronomical PAH model (Li \& Draine 2001b).
\item {\bf PAHs can be rotationally excited} 
by a number of physical processes, 
including collisions with neutral atoms and ions, 
``plasma drag'', and absorption and emission of photons. 
It is shown that these processes can drive PAHs to 
rapidly rotate, with rotation rates 
reaching tens of GHz. The rotational electric dipole 
emission from these spinning PAH molecules is capable 
of accounting for the observed ``anomalous'' microwave 
emission (Draine \& Lazarian 1998a,b; Draine 1999; 
Draine \& Li 2004).
\end{itemize}

\item {\bf (10) \underline{IR Emission from Interstellar Dust.}}
Interstellar grains absorb sta-\\rlight in the UV/visible 
and re-radiate in the IR. The IR emission spectrum of the Milky
Way diffuse ISM, estimated using the IRAS 12, 25, 60 and 100$\mum$
broadband photometry,
the DIRBE-COBE 2.2, 3.5, 4.9, 12, 25, 60, 100, 140 and
240$\mum$ broadband photometry,
and the FIRAS-COBE 110$\mum$$<$$\lambda$$<$3000$\mum$
spectrophotometry, is characterized by a modified
black-body of $\lambda^{-1.7}B_\lambda$(T=19.5$\K$)
peaking at $\simali$130$\mum$ in the wavelength range
of 80$\mum$$\ltsim$$\lambda$$\ltsim$1000$\mum$,
and a substantial amount of emission at $\lambda$$\simlt$60$\mum$
which far exceeds what would be expected from 
dust at $T$$\approx$20$\K$ (see Draine 2003a).
In addition, spectrometers aboard 
the IRTS (Onaka et al.\ 1996; Tanaka et al.\ 1996) 
and ISO (Mattila et al.\ 1996) 
have shown that the diffuse ISM radiates strongly in 
emission features at 3.3, 6.2, 7.7, 8.6, and 11.3$\um$. 
\vskip -0.10in
\begin{itemize}
\vskip -0.15in
\item The emission at $\lambda$$\gtsim$60$\mum$ accounts
for $\simali$65\% of the total emitted power.
$\longrightarrow$ There must exist a population 
of {\bf ``cold dust''} in the size range of
$a$$>$250$\Angstrom$, heated by starlight 
to equilibrium temperatures 15$\K$$\simlt$$T$$\simlt$25$\K$
and cooled by far-IR emission (see Li \& Draine 2001b).
\item The emission at $\lambda$$\ltsim$60$\mum$ accounts
for $\simali$35\% of the total emitted power.
$\longrightarrow$ There must exist a population 
of {\bf ``warm dust''} in the size range of
$a$$<$250$\Angstrom$, stochastically heated by single
starlight photons to temperatures $T$$\gg$20$\K$
and cooled by near- and mid-IR emission
(see Li \& Draine 2001b; Li 2004a).
\end{itemize}

\item {\bf (11) \underline{Interstellar Depletions.}}
Atoms locked up in dust grains are ``depleted'' from 
the gas phase. The dust depletion can be determined 
from comparing the gas-phase abundances measured from
optical and UV absorption spectroscopic lines with the
assumed reference abundances of the ISM 
(total abundances of atoms both in gas and in dust;
also known as ``standard abundances'', 
``interstellar abundances'',  and ``cosmic abundances'').
The total interstellar abundances are usually taken 
to be solar, although Snow \& Witt (1996) argued
that interstellar abundances are appreciably subsolar
($\simali$70\% solar). Interstellar depletions allow us
to extract important information about the composition
and quantity of interstellar dust:
\vskip -0.10in
\begin{itemize}
\vskip -0.15in
\item In low density clouds, Si, Fe, Mg, C, and O are depleted.
$\longrightarrow$ {\bf Interstellar dust must contain an
appreciable amount of Si, Fe, Mg, C and O.} Indeed, all
contemporary interstellar dust models consist of both
silicates and carbonaceous dust. 
\item From the depletion of the major elements Si, Fe, Mg, C, 
and O one can estimate {\bf the gas-to-dust mass ratio to be 
$\simali$165}.\footnote{%
 Let $\xsun$ be the interstellar abundance of X relative to H 
 (we assume interstellar abundances to be those of 
  the solar values: $\csun$$\approx$391 parts per million [ppm],
  $\osun$$\approx$$501\ppm$,
  $\mgsun$$\approx$$34.5\ppm$, $\fesun$$\approx$$34.4\ppm$,
  and $\sisun$$\approx$$28.1\ppm$ [Sofia 2004]);
  $\xgas$ be the amount of X in gas phase
  ($\cgas$$\approx$$130\ppm$, $\ogas$$\approx$$375\ppm$; 
   Fe, Mg and Si are highly depleted in dust:
   $\fegas$$\approx$$1\ppm$, $\mggas$$\approx$$2\ppm$,
  and $\sigas$$\approx$$2\ppm$ [Sofia 2004]);
  $\xdust$ be the amount of X contained in dust
  ($\cdust$$\,=\,$$\csun-\cgas$$\approx$$261\ppm$, 
   $\odust$$\approx$$126\ppm$, $\mgdust$$\approx$$32.5\ppm$,
   $\fedust$$\approx$$27.1\ppm$, $\sidust$$\approx$$32.4\ppm$).
   Assuming H/C=0.5 for interstellar carbon dust, 
   the gas-to-dust mass ratio is
\begin{equation}
 \frac{m_{\rm gas}}{m_{\rm dust}} \approx
 \frac{1.4\,\mu_{\rm H}}{\sum_{\rm X}\xdust\,\mu_{\rm X}}
 \approx 165~.
\end{equation}
where the summation is over Si, Mg, Fe, C, O and H,
and $\mux$ is the atomic weight of X in unit of 
$\muh\approx 1.66\times 10^{-24}\g$.   
}
\item In addition to the silicate dust component, there must 
exist another dust population, since {\bf silicates alone are 
not able to account for the observed amount of extinction 
relative to H} although Si, Mg, and Fe are highly depleted 
in the ISM. Even if all Si, Fe, and Mg elements are locked up 
in submicron-sized silicate grains, they can only account for 
$\simali$60\% of the total observed optical extinction.\footnote{%
 Assuming all Si, Mg, and Fe elements of solar abundances 
 are condensed in silicate dust of a stoichiometric composition 
 MgFeSiO$_4$ with a characteristic size $a$$\approx$0.1$\mum$, 
 the contribution of the silicate dust to the optical extinction is    
\begin{eqnarray}
\nonumber
 \left(\frac{A_V}{N_{\rm H}}\right)_{\rm sil} & \approx &
 1.086\,\pi a^2 Q_{\rm ext}(V) N_{\rm sil}/N_{\rm H}\\
& \approx & \frac{1.086\,\pi a^2 Q_{\rm ext}(V)
\nonumber 
 \left(\sum_{\rm X=Si,Mg,Fe}\xsun\mux\,+\,4\,\sisun\muo\right)
 \muh}{\left(4/3\right)\pi a^3\rho_{\rm sil}}\\
& \approx & 3.2\times 10^{-22}\magni\cm^2~.
\end{eqnarray}
 where $N_{\rm sil}$ is the column density of silicate dust,
 $\rho_{\rm sil}$$\approx$3.5$\g\cm^3$ is the mass density 
 of silicate material, and $Q_{\rm ext}(V)$ is the visual 
 extinction efficiency of submicron-sized silicate dust which 
 is taken to be $Q_{\rm ext}(V)$$\approx$1.5.
}
\end{itemize}
\end{itemize}

\section{\large\hspace{-8mm} Warm and Cold Dust in the ISM}
As summarized in \S2, the interstellar extinction,
polarization, scattering, the near, mid, and far-IR emission
and the 3.3--11.3$\mum$ PAH emission features point
to the existence of two dust populations in interstellar space:
\begin{itemize}
\item There exists a population of large grains 
with $a$$\simgt$250$\Angstrom$. Illuminated by the general
interstellar radiation field (ISRF), these grains, defined 
as {\bf ``\underline{cold dust}''}, 
obtain equilibrium temperatures of 
15$\K$$\simlt$$T$$\simlt$25$\K$ and emit strongly at 
wavelengths $\lambda$$\gtsim$60$\mum$. These grains are
responsible for the near-IR/optical extinction, scattering,
polarization and the emission at $\lambda$$\gtsim$60$\mum$.

The equilibrium temperature $T$ for a large grain of spherical
radius $a$ is determined by balancing absorption and emission:
\begin{equation}\label{eq:calc_Teq}
\int^{\infty}_{0} C_{\rm abs}(a,\lambda) c u_{\lambda} d\lambda
= \int^{\infty}_{0} C_{\rm abs}(a,\lambda) 
  4\pi B_{\lambda}(T)d\lambda~~,
\end{equation}
where $C_{\rm abs}(a,\lambda)$ is the absorption cross section for a grain
with size $a$ at wavelength $\lambda$, $c$ is the speed of light,
$B_{\lambda}(T)$ is the Planck function at temperature $T$, and
$u_{\lambda}$ is the energy density of the radiation field. 
In Figure \ref{fig:temp_equil} we display
these ``equilibrium'' temperatures for graphitic and silicate grains
as a function of size in environments with various UV intensities. 

\begin{figure}[ht]
\vskip -0.25in
\centerline{\includegraphics[width=\figwidth]{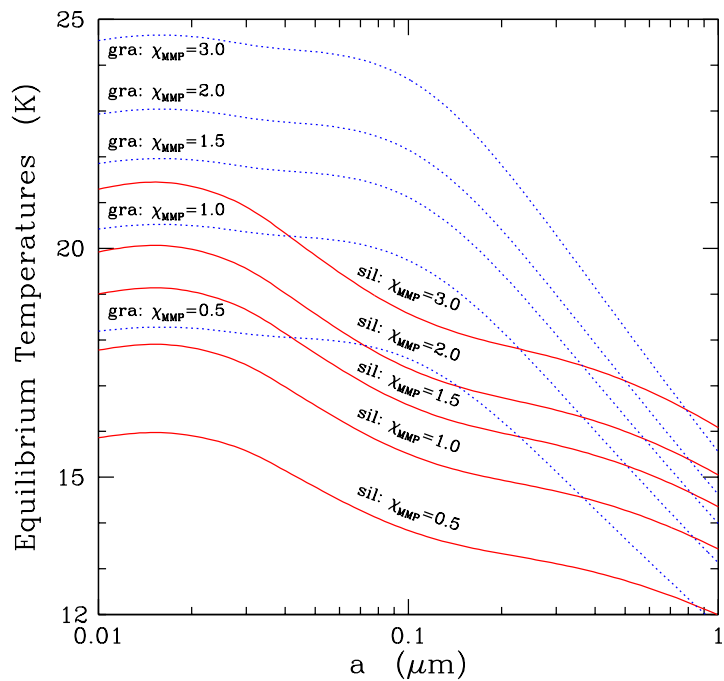}}
\vskip -0.12in
\caption{\footnotesize
        \label{fig:temp_equil}
	Equilibrium temperatures for graphite (dotted lines) 
        and silicate grains (solid lines) in environments with
        various starlight intensities (in units of the Mathis,
        Mezger, \& Panagia 1983 [MMP] solar neighbourhood ISRF).
        Taken from Li \& Draine (2001b).        
        }
\end{figure}
\item There exists a population of ultrasmall grains 
with $a$$\simlt$250$\Angstrom$. These grains have energy
contents smaller or comparable to the energy of a single
starlight photon. As a result, a single photon can heat
a very small grain to a peak temperature much higher than 
its ``steady-state'' temperature and the grain then rapidly
cools down to a temperature below the ``steady-state'' 
temperature before the next photon absorption event.
Stochastic heating by absorption of starlight therefore 
results in transient ``temperature spikes'', during which 
much of the energy deposited by the starlight photon is 
reradiated in the IR -- because of this, we call these
ultrasmall grains {\bf ``\underline{warm dust}''}.\footnote{%
 The idea of transient heating of very small grains was 
 first introduced by Greenberg (1968).
 This process was not observed until many years later
 when the detection of the near-IR emission of reflection
 nebulae (Sellgren, Werner, \& Dinerstein 1983)
 and detection by IRAS of 12 and 25$\mum$ Galactic 
 emission (Boulanger \& P\'erault 1988) were reported.
 }
These grains are responsible for the far-UV extinction rise
and the emission at $\lambda$$\simlt$60$\mum$ (including
the 3.3--11.3$\mum$ PAH emission features), 
dominate the photoelectric heating of interstellar 
gas (see \S2.5 in Li 2004a), and provide most of 
the grain surface area in the diffuse ISM.

Since ultrasmall grains will not maintain ``equilibrium
temperatures'', we need to calculate their temperature
(energy) probability distribution functions in order to
calculate their time-averaged IR emission spectrum.
There have been a number of studies on this topic 
since the pioneering work of Greenberg (1968). 
A recent extensive investigation was carried out by 
Draine \& Li (2001). We will not go into details in 
this review, but just refer those who are interested 
to Draine \& Li (2001) and a recent review article 
of Li (2004a).

\begin{figure}[ht]
\vskip -0.15in
\centerline{\includegraphics[width=\figwidth]{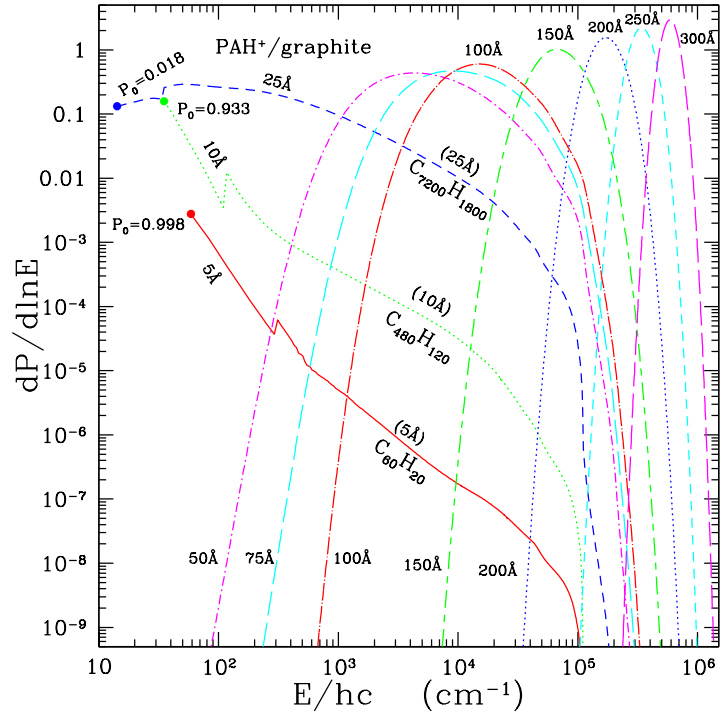}}
\vskip -0.1in
\caption{\footnotesize
        \label{fig:temp_spikes}
        The energy probability distribution functions 
        for charged carbonaceous grains 
        ($a=5\Angstrom$ [C$_{60}$H$_{20}$], 
         10$\Angstrom$ [C$_{480}$H$_{120}$], 
         25$\Angstrom$ [C$_{7200}$H$_{1800}$], 
         50, 75, 100, 150, 200, 250, 300\,\AA)
	illuminated by the general ISRF.
        The discontinuity in the 5, 10, and 25$\Angstrom$ curves 
	is due to the change of the estimate 
        for grain vibrational ``temperature'' at the 20th vibrational
        mode (see Draine \& Li 2001). For 5, 10, and 25\AA\ a dot
        indicates the first excited state, and $P_0$ is
        the probability of being in the ground state.
        Taken from Li \& Draine (2001b).
        }
\end{figure}

For illustration, we show in Figure \ref{fig:temp_spikes} 
the energy probability distribution 
functions $dP/d{\rm ln}E$ (where $dP$ is the probability
that a grain will have vibrational energy in interval
$[E,E+dE]$) for PAHs with radii
$a=5, 10, 25, 50, 75, 100, 150, 200, 300$\AA\ 
illuminated by the general ISRF.
It is seen that very small grains ($a$$\ltsim$$100\Angstrom$) 
have a very broad $P(E)$, and the smallest grains 
($a$$<$$30\Angstrom$) have an appreciable probability $P_0$
of being found in the vibrational ground state $E$=0.
As the grain size increases, $P(E)$ becomes narrower, 
so that it can be approximated by 
a delta function for $a$$>$$250\Angstrom$.\footnote{%
 This is because for large grains individual photon
 absorption events occur relatively frequently and
 the grain energy content is large enough that 
 the temperature increases induced by individual photon
 absorptions are relatively small.
 } 
However, for radii as large as $a$=$200\Angstrom$, 
grains have energy distribution functions 
which are broad enough that the emission spectrum 
deviates noticeably from the emission spectrum for 
grains at a single ``steady-state'' temperature $T$, 
as shown in Figure \ref{fig:irem_spike_ss}. 
For accurate computation of IR emission spectra 
it is therefore important to properly calculate 
the energy distribution function $P(E)$, 
including grain sizes which are large enough that the average 
thermal energy content exceeds a few eV.

\begin{figure}[ht]
\vskip -0.1in
\centerline{\includegraphics[width=\figwidth]{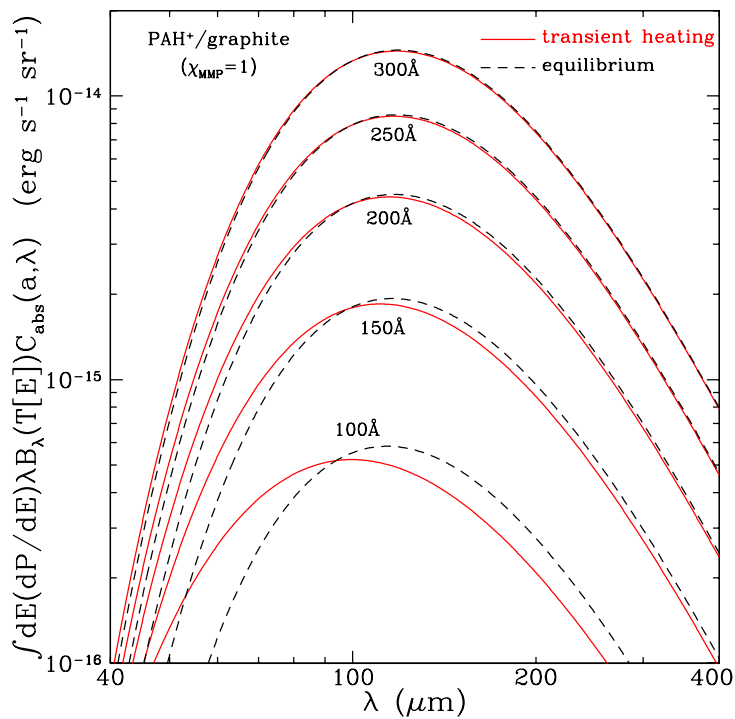}}
\vskip -0.1in
\caption{\footnotesize
        \label{fig:irem_spike_ss}
        Infrared emission spectra for small carbonaceous grains
	of various sizes heated by the general ISRF,
	calculated using the full energy distribution function 
        $P(E)$ (solid lines); also shown (broken lines) are spectra
	computed for grains at the ``equilibrium'' temperature $T$. 
	Transient heating effects lead to
        significantly more short wavelength emission for 
        $a$$\ltsim$200$\Angstrom$. Taken from Li \& Draine (2001b).
        }
\end{figure}

\item Nano-sized {\it interstellar} diamond and TiC grains 
have been identified in primitive meteorites based on 
isotopic anomaly analysis. Exposed to the general ISRF, 
these grains are subject to single-photon heating 
and should of course be classified as ``warm dust''. 
But we should note that they are not representative of 
the bulk interstellar dust (see \S5.4 and \S5.5 in Li 2004a). 
The carriers of the ERE (see \S2.8) and the 2175$\Angstrom$ 
extinction hump (see \S2.4) are also in the single-photon 
heating regime. Therefore, they can also be classified as 
``warm dust''. As a matter of fact, the latter is attributed 
to PAHs (see \S2.4). On the other hand, the Ulysses and Galileo 
spacecrafts have detected a substantial number of 
large interstellar grains with $a$$\simgt$1$\um$
(Gr\"un et al.\ 1994), much higher than expected for 
the average interstellar dust distribution. 
In the ISM, these should of course be considered 
as ``cold dust'' or ``very cold dust'' with $T$$<$10$\K$.
But we note that the reported mass of these large grains
is difficult to reconcile with the interstellar extinction 
and interstellar elemental abundances. 
\end{itemize}

\section{\large\hspace{-8mm} The Silicate-Graphite-PAHs Interstellar Dust Model}
Various models have been proposed for interstellar dust
(see Li \& Greenberg 2003, Li 2004a, Draine 2004 for recent 
reviews). In general, these models fall into three broad 
categories: the silicate core-carbonaceous mantle model
(Li \& Greenberg 1997), the silicate-graphite-PAHs model
(Li \& Draine 2001b, Weingartner \& Draine 2001a) 
and the composite model (Mathis 1996, Zubko, Dwek \& Arendt 2004).
In this review I will focus on the IR emission calculated
from the silicate-graphite-PAHs model and refer those who 
are interested in a detailed comparison between different 
models to my recent review articles 
(Li 2004a, Li \& Greenberg 2003).

The silicate-graphite-PAHs model, consisting of a mixture 
of amorphous silicate dust and carbonaceous dust
-- each with an extended size distribution ranging 
from molecules containing tens of atoms
to large grains $\gtsim$1$\um$ in diameter,
is a natural extension of the classical silicate-graphite 
model (Mathis, Rumpl, \& Nordsieck 1977; Draine \& Lee 1984).
We assume that the carbonaceous grain population extends from 
grains with graphitic properties at radii $a$$\gtsim$50$\Angstrom$, 
down to particles with PAH-like properties at very small sizes.

\begin{figure}[ht]
\vskip -0.1in
\centerline{\includegraphics[width=\figwidthland,angle=270]{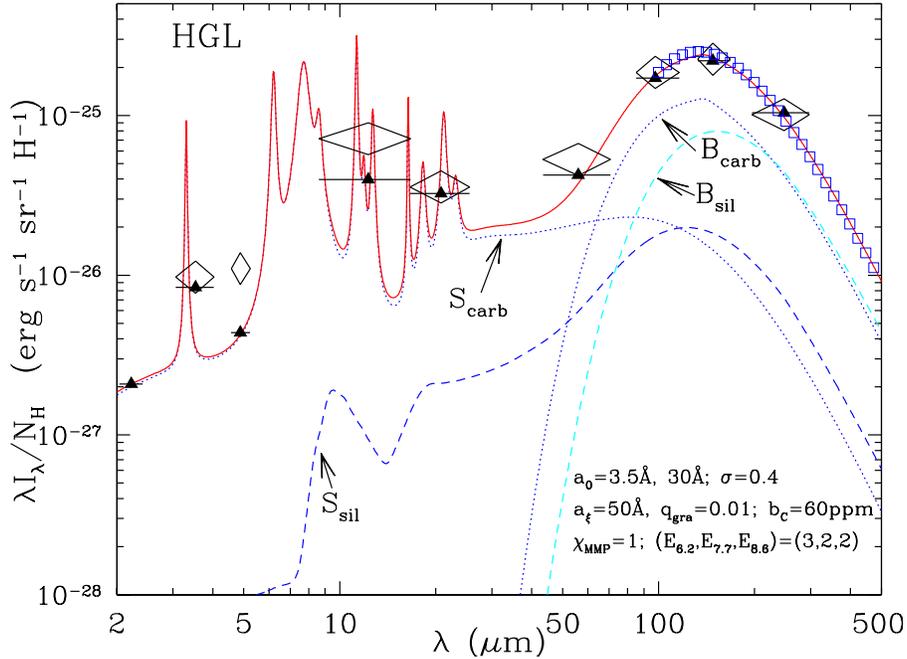}}
\vskip -0.1in
\caption{
	\footnotesize
        \label{fig:hgl}
        Comparison of the model to the observed
        emission from the diffuse ISM at
        high galactic latitudes ($|b|$$\ge$25$^{\rm o}$).
	Curves labelled B$_{\rm sil}$ and B$_{\rm carb}$ show emission
	from ``big'' ($a$$\geq$250$\Angstrom$) silicate and carbonaceous 
	grains; curves labelled S$_{\rm sil}$ and S$_{\rm carb}$ show
	emission from ``small'' ($a$$<$250$\Angstrom$) silicate and
	carbonaceous grains (including PAHs). 
        Triangles show the model spectrum (solid curve)
	convolved with the DIRBE filters.
        Observational data are from 
        DIRBE (diamonds) and FIRAS (squares).
        Taken from Li \& Draine (2001b).
        }
\end{figure}

\begin{figure}[ht]
\vskip -0.1in
\centerline{\includegraphics[width=\figwidthL]{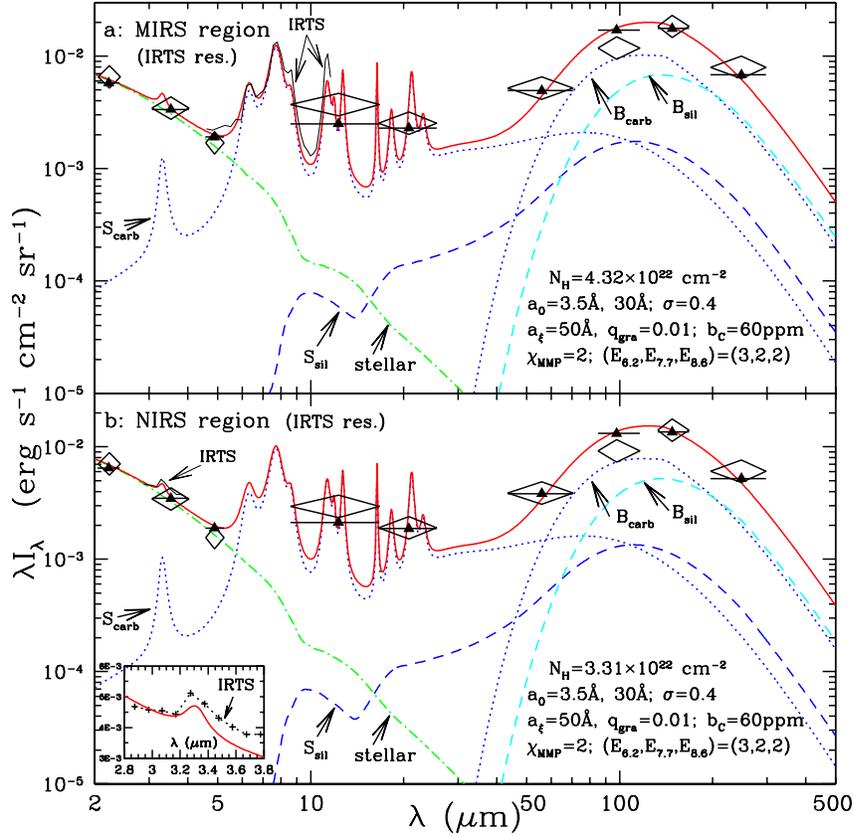}}
\vskip -0.1in
\caption{
	\footnotesize
        \label{fig:plane}
	Infrared emission from dust plus 
	starlight for two regions
	in the Galactic plane: (a) the MIRS region
        ($44^{\rm o}\le l \le 44^{\rm o}40^\prime$, 
        $-0^{\rm o}40^\prime\le b \le 0^{\rm o}$), and
	(b) the NIRS region
	($47^{\rm o}30^\prime\le l \le 48^{\rm o}$, $|b|\le 15^\prime$).
	The starlight intensity heating the dust 
	has been taken to be twice the MMP ISRF.
	The solid curve shows the overall model spectrum;
        triangles show the model 
        spectrum convolved with the DIRBE filters.
	DIRBE observations are shown as diamonds.
	For the MIRS field we show the IRTS MIRS 5--12$\micron$ spectrum 
	(thin solid line).
	For the NIRS field we show the IRTS NIRS 2.8-3.9$\micron$ spectrum
	(thin solid line, also shown as cross-dotted curve in inset).
        Taken from Li \& Draine (2001b).
        }
\end{figure}

With the temperature (energy) probability distribution
functions calculated for ultrasmall grains undergoing 
``temperature spikes'' and equilibrium temperatures
calculated for large grains illuminated by the local ISRF, 
the silicate-graphite-PAHs model with grain size distributions 
consistent with the observed $R_V$=3.1 interstellar extinction 
(Weingartner \& Draine 2001a), is able to reproduce 
the observed near-IR to submillimeter emission spectrum
of the diffuse ISM, including the PAH emission features 
at 3.3, 6.2, 7.7, 8.6, and 11.3$\micron$. 
This is demonstrated in Figure \ref{fig:hgl}
and Figure \ref{fig:plane} for the high-latitude ``cirrus''
cloud and 2 regions in the Galactic plane
(see Li \& Draine 2001b for details).

\begin{figure}[ht]
\vskip -0.1in
\centerline{\includegraphics[width=\figwidthland,angle=270]{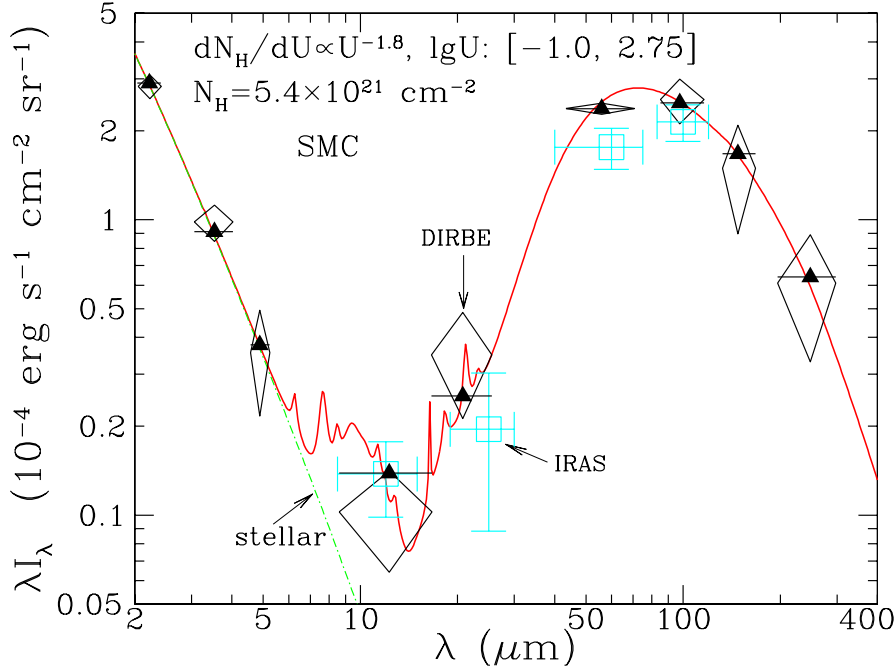}}
\vskip -0.1in
\caption{
	\footnotesize
        \label{fig:smc}
	Comparison of the model (solid line) to the observed
        emission from the SMC obtained by COBE/DIRBE (diamonds)
        and IRAS (squares) averaged over a 6.25 deg$^2$ region
        including the optical bar and the Eastern Wing.
        Triangles show the model spectrum 
        convolved with the DIRBE filters.
        Stellar radiation (dot-dashed line) dominates 
        for $\lambda$$\simlt$6$\mum$. 
        Grains are illuminated by a range of
        radiation intensities $\dN/d\Urad\propto \Urad^{-1.8}$,
        0.1$\le$$\Urad$$\le$$10^{2.75}$,  
        with $N_{\rm H}^{\rm tot}\approx 5.4\times 10^{21}\cm^{-2}$.
        Taken from Li \& Draine (2002c).
        }
\end{figure}

The silicate-graphite-PAHs model, with size distributions
consistent with the SMC Bar extinction curve 
(Weingartner \& Draine 2001a), is also successful 
in reproducing the observed IR emission from 
the SMC (Li \& Draine 2002c), as shown in Figure \ref{fig:smc}.
The dust in the SMC is taken to be illuminated by 
a distribution of starlight intensities. 
Following Dale et al.\ (2001), we adopt a simple 
power-law function for the starlight intensity distribution.
The SMC, with a low metallicity ($\simali$10\% of solar)
and a low dust-to-gas ratio ($\simali$10\% of the Milky Way),
has a very weak or no 2175$\Angstrom$ extinction hump
in its extinction curves for most sightlines (see \S2.1)
and very weak 12$\mum$ emission (see Fig.\,\ref{fig:smc})
which is generally attributed to PAHs, 
supporting the idea of PAHs as the carrier for 
the 2175$\Angstrom$ extinction hump.\footnote{%
  Li \& Draine (2002c) placed an upper limit of
  $\simali$0.4\% of the SMC C abundance on the amount 
  of PAHs in the SMC Bar. But we note that the PAH emission 
  features have been seen in SMC B1\#1, a quiescent molecular 
  cloud (Reach et al.\ 2000). 
  For this region, Li \& Draine (2002c) estimated that
  $\sim$3\% of the SMC C abundance to be incorporated into PAHs.
  }

\begin{figure}[ht]
\vskip -0.1in
\centerline{\includegraphics[width=\figwidthL]{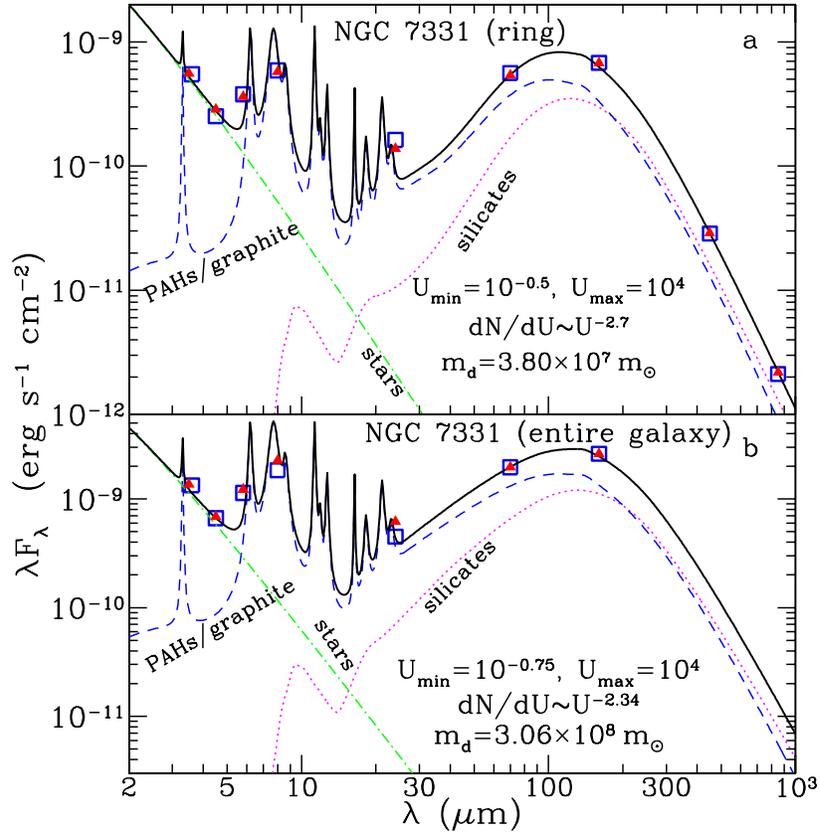}}
\vskip -0.1in
\caption{
	\footnotesize
        \label{fig:ngc7331}
        IR emission and model fits to the NGC\,7331 ring (a) and
        the entire galaxy (b). The thick solid lines 
        and triangles are the model-predicted fluxes, 
        and the squares are the observed fluxes. 
        The broken curves indicate the contributions 
        of the different model components. 
        Taken from Regan et al.\ (2004).
        }
\end{figure}

Very recently, the silicate-graphite-PAHs model has also
been successfully applied to NGC\,7331, a ringed Sb galaxy.  
Using the same set of dust parameters determined for the Milky
Way diffuse ISM ($R_V$=3.1; Weingartner \& Draine 2001a),
as shown in Figure \ref{fig:ngc7331},
this model fits the IR emission observed by 
the IRAC instrument at 3.6, 4.5, 5.8 and 8$\mum$ 
and the MIPS instrument at 24, 70 and 160$\mum$
aboard the Spitzer Space Telescope 
and the 450 and 850$\mum$ SCUBA
submillimeter emission observed by JCMT,
both for the ring and inside star-forming region
and for the galaxy as whole 
(see Regan et al.\ 2004 for details).
The model also closely reproduces the observed 
6.2, 7.7, 8.6, 11.3 and 12.7$\mum$ PAH emission
features (see Fig.\,2 of Smith et al.\ 2004). 

\section{\large\hspace{-8mm} Very Cold Dust or How Cold Could Galaxies Be?}
In the 1996 South African ``{\it Cold Dust}'' Symposium
(Block \& Greenberg 1996), the possible existence 
of a population of {\bf very cold dust}
(with equilibrium temperatures $T$$<$10$\K$) in 
interstellar space was a subject received much attention.
In his invited paper titled ``{\it How Cold Could Galaxies Be?}''
published in the proceedings for that symposium, Mike Disney
wrote ``... {\it An eminent cosmologist once advised me to forget 
all about very cold dust because the $T^4$ law ensures that 
it cannot emit, and therefore by implication cannot absorb, 
much radiation. He sounded plausible, as Cosmologists are apt 
to sound, but he was in fact totally wrong, as Cosmologists 
are apt to be.}'' 

Disney (1996) argued that for grains with higher IR and 
far-IR emissivities, they can achieve rather low temperatures. 
At first glance, this appears plausible as can be seen in 
Eq.(\ref{eq:calc_Teq}): for a given interstellar radiation 
field, grains with fixed UV and optical absorption properties
would obtain lower temperatures if their long wavelength 
emissivities are enhanced. Therefore, Disney (1996) wrote 
``... {\it [Since] we are not confident about their 
[interstellar grains] size distribution 
and their emissivities, particularly at long wavelength, 
... {\bf we have to keep our minds open to the possible 
existence of a significant amounts of very cold ($<$10$\K$) 
material in spiral galaxies.}}'' 

So far, the detection of very cold dust 
($T$$<$10$\K$) in interstellar space on 
galactic scales has been reported for various objects:
NGC\,4631, a low metallicity ($\simali$1/2 of solar) 
interacting galaxy with $T$$\simali$4--6$\K$ 
(Dumke, Krause, \& Wielebinski 2004);
NGC\,1569, a low metallicity ($\simali$1/4 of solar) 
starbursting dwarf galaxy with 
$T$$\simali$5--7$\K$ (Galliano et al.\ 2003);
inactive spiral galaxies UGC\,3490 with 
$T$$\simali$9$\K$ (Chini et al.\ 1995),
NGC\,6156 with $T$$\simali$8.6$\K$,
and NGC\,6918 with $T$$\simali$9.4$\K$ 
(Kr\"ugel et al.\ 1998);
and several irregular and blue compact dwarf galaxies 
with $T$$<$10$\K$ in the Virgo Cluster
(Popescu et al.\ 2002).
Very cold dust with $T$$\simali$4--7$\K$
has also been detected in the Milky Way Galaxy
(Reach et al.\ 1995; also see Boulanger et al.\ 2002).
This component is widespread and spatially correlated
with the warm component (16--21$\K$).
By comparing the dust mass calculated from
the IRAS data with the molecular and atomic gas
masses of 58 spiral galaxies, Devereux \& Young (1990)
argued that the bulk of the dust in spiral galaxies 
is $<$15$\K$ regardless of the phase of the ISM. 

How can dust get so cold? In literature, suggested 
solutions include (1) the dust is deeply embedded in 
clumpy clouds and heated by the far-IR emission from 
``classical grains'' (Galliano et al.\ 2003; 
Dumke et al.\ 2004);\footnote{%
 Siebenmorgen et al.\ (1999) argued that
 the dust embedded in UV-attenuated clouds within 
 the optical disk of typical inactive spiral galaxies 
 cannot become colder than $\simali$6$\K$.
 } 
(2) the dust has unusual optical properties (e.g. fractal or porous 
grains with enhanced submm and mm emissivity;
Reach et al.\ 1995; Dumke et al. 2004). 
{\bf However, as discussed in detail by Li (2004b), 
while the former solution appears to be inconsistent
with the fact that the very cold dust is observed on
galactic scales, the latter violates the Kramers-Kronig
dispersion relation (Purcell 1969; Draine 2003b; Li 2003b),}
except for extremely elongated conducting dust (Li 2003a).
Perhaps the submm and mm excess emission (usually attributed 
to very cold dust) is from something else? 
To avoid the ``temperature'' problem,
one can adopt the Block direct method to identify
and characterize the presence and distribution of 
the cold and very cold dust:
using near-IR camera arrays and subtracting these 
from optical CCD images. This method measures the dust extinction 
cross section and does not require the knowledge of dust 
temperature (see Block 1996; Block et al.\ 1994a,b, 1999).
Otherwise, a detailed radiative transfer treatment
of the interaction of the dust with starlight
(e.g. Popescu et al.\ 2000, Tuffs et al.\ 2004)
together with a physical interstellar dust model
(e.g. Li \& Draine 2001b,2002c) is required.

\section{\large\hspace{-8mm} Can Dust Get Down to 2.7$\K$ 
and Appear in Absorption against the CMB?}
{\noindent Ultrasmall grains spend most of their time at 
their vibrational ground state during the interval 
of two photon absorption events 
(\S3 and Fig.\,\ref{fig:temp_spikes}; Draine \& Li 2001).   
In the 1996 South African ``{\it Cold Dust}'' Symposium,
it was heatedly argued that these grains could obtain
a vibrational temperature less than the 2.7$\K$ temperature
of the CMB so that they could be detected in absorption 
against the CMB (Duley \& Poole 1998). 
However, based on detailed modeling of the excitation 
and de-excitation of these grains, we found that
even though these grains do have a large population in
the vibrational ground state, nevertheless the vibrational
levels are sufficiently excited that the grains would appear 
in emission against the CMB with brightness temperature
$\simlt$9$\K$ (see Draine \& Li 2004b for details).}\\

{\small\noindent {\bf Acknowledgments}~~ I thank
G.J. Bendo, D.L. Block, F. Boulanger, B.T. Draine,
R.C. Kennicutt, J.I. Lunine, D. Pfenniger, J.L. Puget, 
M.W. Regan, and C. Yuan for helpful discussions.
I am grateful to D.L. Block and the organizing committee
for inviting me to this stimulating symposium.
I also thank my advisors B.T. Draine and 
the late J.M. Greenberg for guiding me to 
this fascinating field -- cosmic dust.}

\end{document}